# TIME TRANSFER THROUGH OPTICAL FIBERS (TTTOF): FIRST RESULTS OF CALIBRATED CLOCK COMPARISONS

**Dirk Piester[1], Miho Fujieda[2], Michael Rost[1], Andreas Bauch[1]**

[1]**Physikalisch-Technische Bundesanstalt (PTB)
Bundesallee 100, 38116 Braunschweig, Germany
E-mail:** *dirk.piester@ptb.de*

[2]**National Institute of Information and Communications Technology (NICT)
Tokyo, Japan**

### Abstract

*We have developed a means for accurate time transfer using optical fibers and aim at the synchronization of clocks located at different places on an institute campus with an overall uncertainty of 100 ps or better. Such an installation shall be used as a part of the infrastructure connecting the ground station setups during forthcoming T2L2 and ACES experiments and the local installations at the PTB time laboratory. Our target transmission length is less than 1 km. To transfer time, a code-domain-multiple-access (CDMA) signal is used for modulation of a laser. Optical signals are exchanged in the two-way mode to cancel long-term fiber length variation. This is similar to the well known two-way satellite time and frequency transfer (TWSTFT) scheme. We discuss procedures for a proper calibration of such time transfer through optical fibers links and show first promising results of an experiment using a test loop on the PTB campus with a length of 2 km.*

## 1. INTRODUCTION

During the last years the transfer of stabilized optical carrier frequency signals through optical fibers has shown ultra low instabilities on distances up to more than 100 km [1]. Very promising results have also been reported for frequency transfer of standard 10 MHz signals [2] and of a 1.5 GHz signal for radio astronomy applications modulated on optical carriers [3]. Phase noise accumulated along the transmission path was cancelled by using a round-trip signal in all examples. The use of optical fibers has the advantage to transmit a signal with less loss and better isolation from electrical noise, resulting in a better transfer stability than feasible with rf-cables.

As an extension to previous set-ups, we have developed a means for accurate time transfer using optical fibers. We aim at the synchronization of clocks located at different places on PTB institute campus. Such an installation shall become part of the infrastructure connecting the ground station setups during the forthcoming T2L2 [4] and ACES [5] experiments with the local installations of the time laboratory at the Physikalisch-Technische Bundesanstalt (PTB). Additionally it was decided to move the PTB TWSTFT ground stations, which are currently spread over the PTB campus, to a common location at a new site (see Figure 1 and [6] for details). From late 2010 onwards the TWSTFT stations will be installed on top of a high building, where free sight to all directions is available. It is also planned to set up the ACES ground terminals at this location, when it becomes available. Especially for this project it is necessary to estab-lish a time scale in an extremely well known relation to UTC(PTB) in a "satellite" time laboratory next to the time transfer equipment for calibration and monitoring issues.

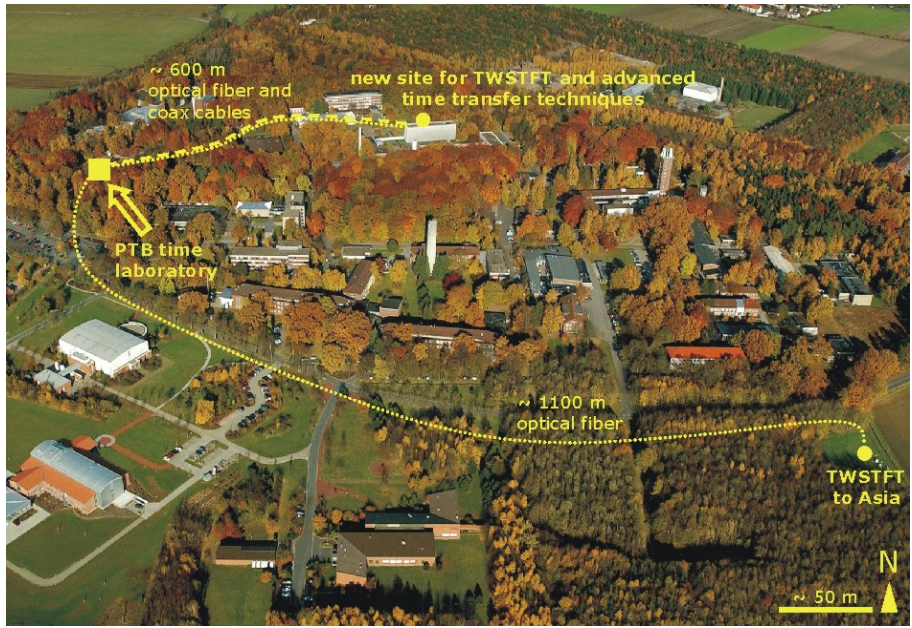

Figure 1: Aerial view of the Braunschweig PTB campus, current optical fiber connection to the ground station for TWSTFT to Asia, and future installations to connect the new site for TWSTFT and advanced time transfer techniques with the PTB time laboratory.

Our target transmission length is below 1 km. Some of the optical signals are exchanged in the two-way mode to cancel long-term fiber length variations [7]. Such an approach was initially proposed by Kihara and Imaoka [8] in the framework of network synchronization. Our time transfer through optical fibers (TTTOF) method is similar to the well known two-way satellite time and frequency transfer (TWSTFT) scheme which already has proven its feasibility for high accuracy calibrated time transfer [9]. In both cases, code-domain-multiple-access (CDMA) signals are used, in TTTOF for the modulation of the laser current.

In the following Section 2 we introduce the basic concept and the experimental setup. Thereafter, in Section 3 the time transfer stability of the used system is analysed in detail. In Section 4 we discuss the possibility for time transfer via opical fibers and report on testing the independence of the time transfer results from the optical fiber length. A summary and an outlook to future experiments will be given in Section 5.

## 2. CONCEPT AND SETUP

The basic concept for the installations in the remote "satellite" time laboratory includes the generation and distribution of a reference frequency (10 MHz) and time scale (1PPS) at the remote site. Both kinds of signals should be related to the reference clock at PTB's local time laboratory. In Figure 2 the basic setup is depicted. The reference frequency is modulated on an optical carrier signal by an electro-optical converter (E/O) and transferred to the remote setup through a single mode optical fiber. At the remote site, it is transformed and distributed by an opto-electrical converter (O/E) and frequency distribution amplifier (FDA), respectively. We use Ortel 10382S Fiberoptic Transmitter and 10481S Fiberoptic Re-ceiver for this purpose. The laser wavelength is nominally 1310 nm. One output of the FDA feeds a divider (DIV) to provide 1PPS signals and the connected pulse distribution amplifier (PDA). One 1PPS output of the PDA is defined as the remote reference time scale TA(2). Initially, TA(2) has to be synchronized to TA(1) by a portable clock or by other suitable means. In principle, the relation between TA(2) and TA(1) is fixed, but affected by the delay instabilities on the transfer path. The delay change due to temperature variation in a spooled 1 km optical fiber, e.g., was reported to be about 30 ps/K/km [10]. The instability introduced by the one-way frequency transfer over an optical fiber of about only 1 km, however, is small enough for our target, so that we omitted for now any phase correction systems in order to ensure a simple and reliable operational setup.

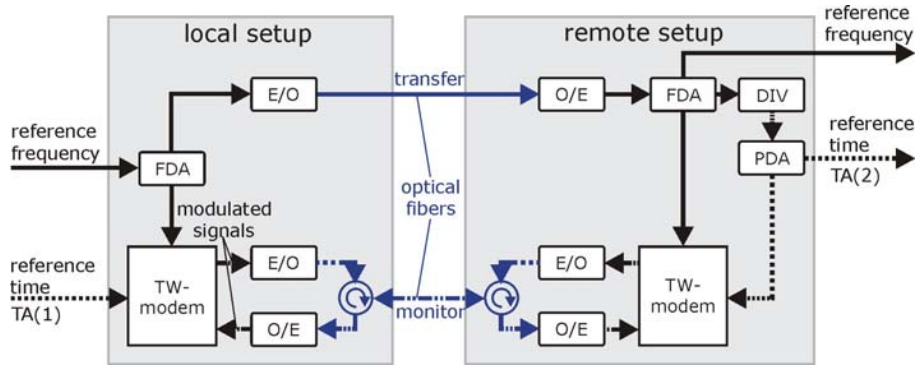

Figure 2: The setup: A remote time scale is generated following a one-way frequency transfer. The time scale is monitored by a two-way measurement system employing modems usually used for TWSTFT (for details see the text).

Instead we measure the difference between TA(2) and TA(1) using TWSTFT modems (type TimeTech SATRE) which are widely used in many time laboratories. In operation the modems at each site exchange binary phase-shift keyed (BPSK) signals. Each modem transmits a 70 MHz signal modulated with a 20 MCh/s BPSK sequence phase coherent to its reference TA($i$). In both modems the phase offset between the signal sent from the counterpart modem and the local reference is measured by the cross correlation method. From these measurements the difference TA(1) – TA(2) can be computed after suitable calibration of the equipment internal delays. The signals from the modems are transformed by using E/Os and O/Es of the same type as mentioned above, but with laser wavelengths of nominally 1550 nm. Additionally we used standard telecommunication circulators and isolators. For clarity the isolators and also additional electrical amplifiers or attenuators are not displayed in the figures throughout this paper.

We use additional experimental equipment for the characterization of the time and frequency transfer. In Section 3 the instability of both the one-way frequency transfer as well as the two-way time transfer are characterized by using a phase comparator type VREMYA-CH VCH-312 and for tests a time interval counter of type SRS SR620. The measurement result of the two-way time transfer should be independent of the length of the optical fiber between the two circulators. As discussed in Section 4, we are using different fiber lengths in a laboratory experiment, i.e. a test loop using a short 15 m fiber and a loop of overall 2.2 km length on the PTB campus (see Figure 1) to verify this assumption.

## 3. FREQUENCY INSTABILITY

The instability of the transfer system was characterized using the laboratory setup as depicted in Figure 3. Here we used an optical fiber link of 2 km buried in the PTB campus for the measurements (see Figure 1). A reference frequency from hydrogen maser H5 (PTB ID) is transferred from the "local" FDA (left hand in Figure 3) to the "remote" one (right hand in Figure 3) via the 2 km optical fiber test loop. The phase comparator is connected directly to both FDAs and measures the phase difference between the two frequency signals. The two-way measurement hardware needs DIVs after both FDAs to feed the modems with 1PPS signals. Here the time difference between the two 1PPS signals is measured by the modems. We used a second 2 km fiber of the same buried cable as above for the two-way measurement system. A test length of 2 km allows a characterization of the future setup in a realistic scenario.

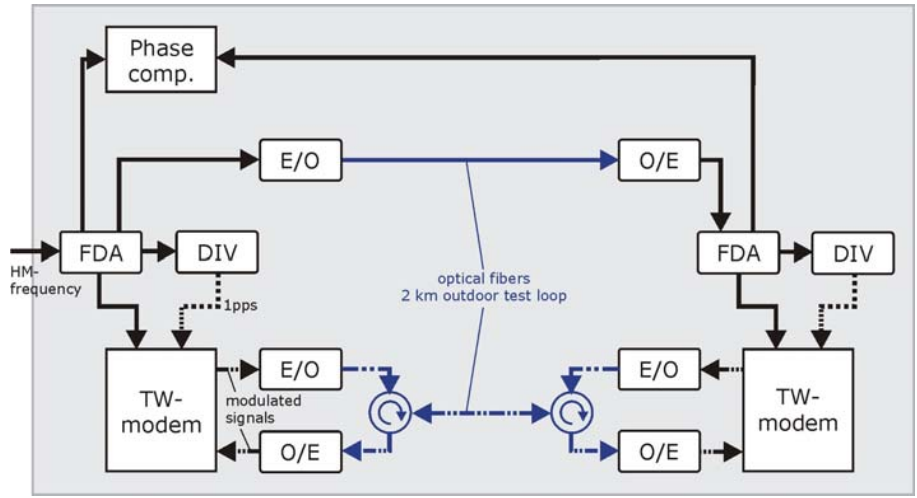

Figure 3: Setup to measure the instability of the one-way fiber frequency transfer as well as the instability of the two-way monitor measurements (both via the 2 km optical fiber outdoor test loop) by means of a phase comparator.

In Figure 4 the measurement results are depicted. Both measurements (two-way and phase comparator) show similar results. The 1-s-data two-way measurements exhibit significant phase noise due to the modems, which can be effectively reduced by applying a 5 minutes averaging moving window to the data. The same process is applied to the phase comparator data. Linear fits to both data sets have slopes of $-5.12 \cdot 10^{-16}$ and $-5.26 \cdot 10^{-16}$, respectively, and thus agree well within $2 \cdot 10^{-17}$. We attribute the observed delay drift to a variation of the optical length of the test fiber loop. In a further step the residual phases of the 5-min-data from the linear fits are computed (Figure 4 bottom). Both data sets show similar long term variations. The double difference, suppressing all common mode instabilities of the one-way frequency transfer, i.e. those caused by FDAs, E/O, fiber, and O/E, shows good agreement after about MJD 55113.7. After this epoch, the overall phase variation is less than 30 ps.

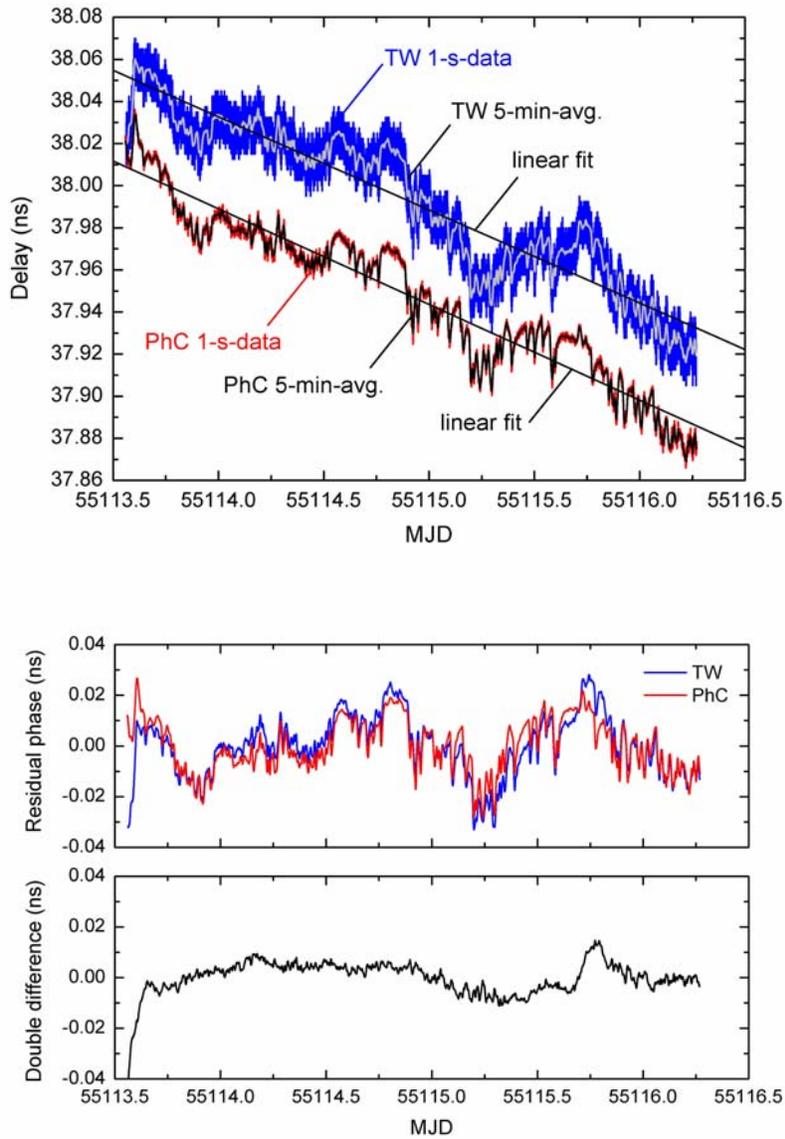

Figure 4: Top: Phase measurements between 10 MHz reference and the frequency transmitted through the 2 km optical fiber test loop: Two-way optical fiber time transfer (blue) and phase comparator measurements (red). Bottom: Residuals to linear fits to the 5-min moving averages, of the two techniques (upper graph) and double difference between both (lower graph).

The frequency instability in terms of the modified Allan deviation are depicted in Figure 5. The instability of the two-way transfer is $6·10^{-12}$ at an averaging time of 1 s. This is in good agreement with the noise level one would expect for this type of modem operated at 20 MCh/s BPSK [11]. Phase noise is dominant until averaging times of about 100 s. The phase comparator measurements have significant lower phase noise at 1 s averaging time but show the same excess noise at averaging times of about 500 s as the two-way data. In the bottom plot of Figure 5, the measurement results using a short fiber of 5 m are de-picted for comparison. They should represent the system performance of the one-way fiber transfer and of the two-way monitor systems. We can thus attribute the excess instability around a few hundreds seconds in the left plot to the 2 km optical fiber test loop. The modified Allan deviation of the double difference is significantly lower than both the two-way and phase comparator data. Having in mind the cancellation of the common-mode noise, the double difference represents the instability due to the uncommon equipment and it agrees well with the system performances shown in the lower plot of Figure 5.

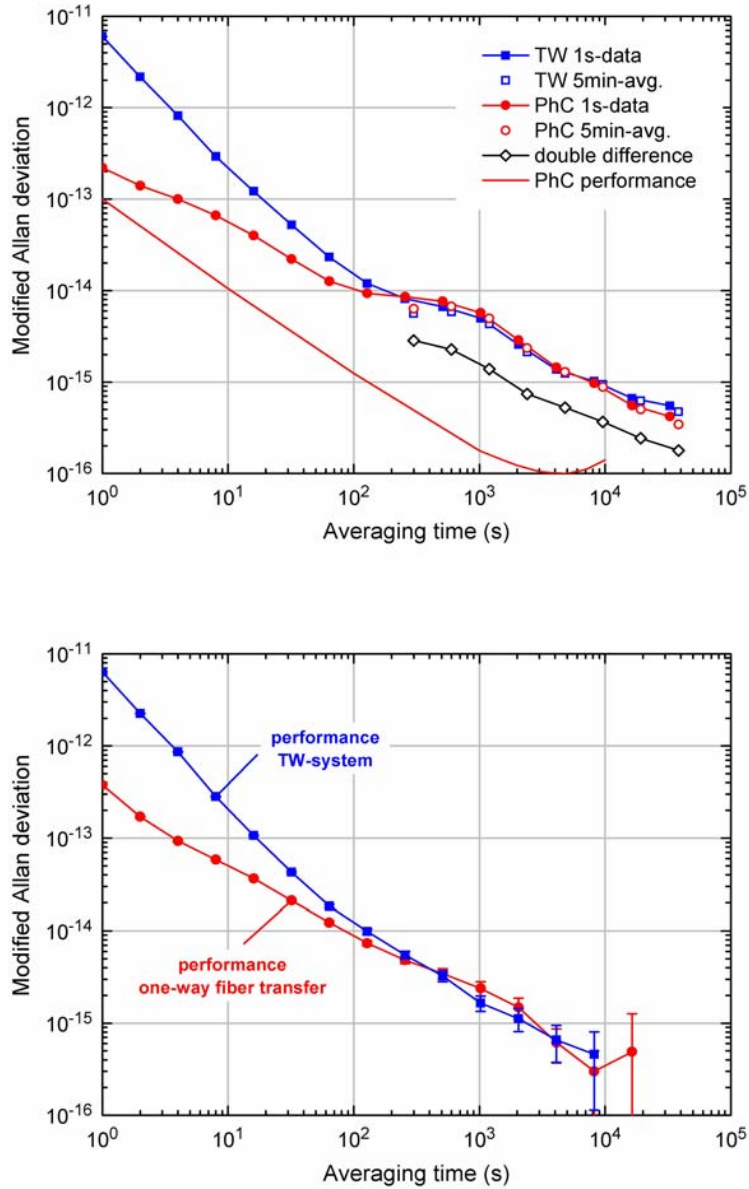

Figure 5: Top: Fractional frequency instability of the two-way measurements (blue squares), phase comparator measurements (red cycles), and double differences (black diamonds). The phase comparator performance is depicted as a red line. Bottom: System performances measured using a short fiber.

As a summary, one-way frequency transfer via the 2 km test loop optical fiber is possible at the $10^{-15}$ level after $10^4$ s averaging and the two-way monitoring system shows an instability well below that value.

## 4. TIME ACCURACY

A prerequisite for accurate comparisons of remote time scales is the possibility for delay calibration of the whole system. Because the optical fiber length of the final setup is unknown, a calibration test is needed to ensure the independence of the setup from the length of the used fiber. For this purpose we connected the modems to reference frequency and 1PPS from FDAs and DIVs as illustrated in Figure 6, and changed the length of the fibers. We chose two different fiber lengths (a 15 m indoor fiber and the 2 km outdoor loop) and additionally we performed a switch off and on procedure of all involved equipment to simulate the anticipated transfer of the setup to the remote location. The attenuation of the two optical fibers was chosen to be at the same level to minimize the impact of receive power dependent delay variations in the modems.

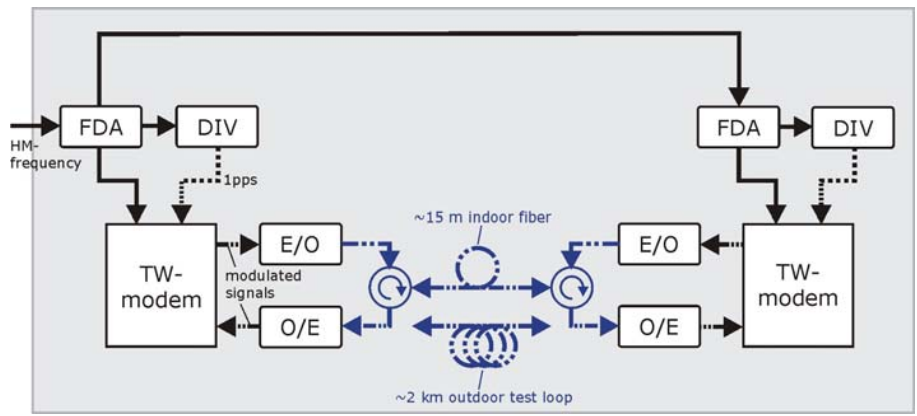

Figure 6: Calibration setup for testing the independence of the time transfer results from the length of the optical fiber. Two lengths were used: a 15 m indoor fiber and the 2 km outdoor test loop.

The schedule we used is as follows: 1) operation with 15 m fiber for 16 h, 2) operation with 2 km fiber for 19 h, 3) equipment power off for 2 h, and 4) operation with 15 m fiber for 40 h. The receive power in both modems was kept constant within ± 0.8 dB. The results are depicted in Figure 7. A delay change of less than 40 ps has been observed between the first and the last measurement. However, the observed variation between the first and the last measurement is significantly larger than the statistical uncertainty represented by the standard deviation of the 1-s measurements. It is not clear presently if these variations are due to power dependent delay variations in the receive path of the modems or due to other effects. In the lower graph of Figure 7 the two-way results after the equipment power off – power on sequence are shown. We attribute the apparent strong drift in the beginning of the data recordings to the heating up process in the hardware. But it also suggests a more detailed investigation of equipment characteristics. Nevertheless, a variation of less than 40 ps under different experimental conditions is a promising first result and well below the aim of 100 ps.

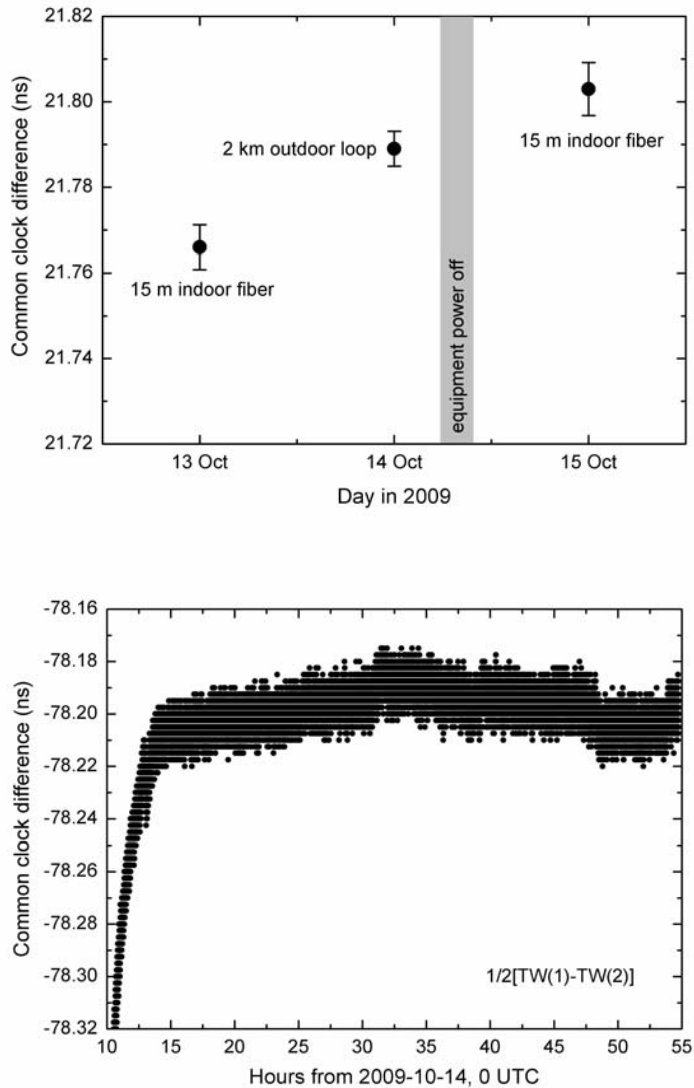

Figure 7: Top: Display of the sequence of calibration measurements results. Bottom: Phase drift observed in the two-way measurements after switching the equipment power off and on (a -100 ns systematic offset is not removed here).

## 5. SUMMARY

We have developed a method to monitor a remote time scale by means of a two-way optical fiber time transfer system. This method is combined with a one way fiber transfer of the reference frequency to generate the time scale at the remote site with sufficient instability. Because our goal is to establish an institute campus solution with distances of about 1 km we have not introduced any phase variation compensation system but rely on monitoring of the fiber related instabilities. We have shown that the one-way frequency transfer via an optical fiber length of 2 km is possible at the $10^{-15}$ level after $10^4$ s averaging. The two-way monitoring system shows an instability well below $10^{-15}$ after $10^4$ s averaging. For accurate time transfer we tested the independence of the time transfer results from the transmission path. A variation of less than 40 ps under different experimental conditions has been observed and is a promising first result. The apparent drift, however, has to be studied in more detail. Especially the receive power dependence of the modems' internal delays seems to be a crucial point and needs more detailed investigations. On the other hand, the operational system will bridge a distance of only 600 m. So the observed instabilities attributed to the 2 km optical fiber should represent an upper limit of instabilities to be expected.

Beside this, future investigations will address the long term validation of the time transfer accuracy, as well as other concepts for the two-way monitor scheme, e.g. different wavelengths of the lasers in the two-way monitor system [7]. Also phase compensation systems, like a phase shifter or fiber heater [12], derived from the real time solutions delivered by the SATRE modems' data outputs could help to improve the stability of the one-way frequency transfer.

The general goal for optical fiber connections as described in this paper are the support of advanced satellite frequency and time transfer techniques as TWSTFT carrier phase or the forthcoming T2L2 and ACES experiments. Thus also an evaluation of the short term instability is required, and last but not least a verification of interface compliance to these experiments has to be achieved.

## 6. ACKNOWLEDGEMENT

The authors wish to thank Wolfgang Schäfer (TimeTech GmbH) for the loan of one SATRE modem and helpful discussions.

## DISCLAIMER

The Physikalisch-Technische Bundesanstalt as a matter of policy does not endorse any commercial product. The mentioning of brands and individual models seems justified here, because all information provided is based on publicly available material or data taken at PTB and it will help the reader to make comparisons with own observations.